\documentstyle[12pt]{article}
     \def\lsim{\raise0.3ex\hbox{$<$\kern-0.75em\raise-1.1ex\hbox{$\sim$}}}
\def\gsim{\raise0.3ex\hbox{$>$\kern-0.75em\raise-1.1ex\hbox{$\sim$}}}
\def\noi{\noindent}
\def\nn{\nonumber}
\def\bea{\begin{eqnarray}}  \def\eea{\end{eqnarray}}
\def\beq{\begin{equation}}   \def\eeq{\end{equation}}

\def\beeq{\begin{eqnarray}} \def\eeeq{\end{eqnarray}}
\def\R{ {\rm R \kern -.31cm I \kern .15cm}}
\def\C{ {\rm C \kern -.15cm \vrule width.5pt \kern .12cm}}
\def\Z{ {\rm Z \kern -.27cm \angle \kern .02cm}}
\def\N{ {\rm N \kern -.26cm \vrule width.4pt \kern .10cm}}
\def\1{{\rm 1\mskip-4.5mu l} }

\begin{document}
\begin{center}
{\large \bf Bounds on the derivatives of the Isgur-Wise function} \\
\vspace{3 truemm}
{\large \bf from sum rules in the heavy quark limit of QCD}\\
\vskip 1 truecm
{\bf A. Le Yaouanc, L. Oliver and J.-C. Raynal} \par \vskip 5 truemm
{\it Laboratoire de Physique Th\'eorique}\footnote{Unit\'e Mixte de Recherche
UMR 8627 - CNRS }\\    {\it Universit\'e de Paris XI, B\^atiment 210, 91405
Orsay Cedex, France}
\end{center}

\begin{abstract}
Using the OPE and the trace formalism, we have obtained a number of 
sum rules in the heavy quark limit of QCD that include the sum over 
all excited states for any
value $j^P$ of the light cloud. We show that these sum rules imply 
that the elastic Isgur-Wise
function $\xi (w)$ is an alternate series in powers of $(w-1)$. 
Moreover, one gets that any $n$-th derivative of $\xi (w)$ at $ w=1$ 
can be bounded by the $(n-1)$-th
one. For the curvature $\sigma^2 = \xi''(1)$, this proves the already 
proposed bound $\sigma^2 \geq {5 \over 4} \rho^2$. Moreover, we 
obtain the absolute lower
bound for the $n$-th derivative $(-1)^n \xi^{(n)}(1) \geq {(2n+1)!! 
\over 2^{2n}}$, that generalizes the results $\rho^2 \geq {3 \over 
4}$ and $\sigma^2 \geq {15
\over 16}$. \end{abstract}

\vskip 2 truecm

\noi LPT Orsay 02-94 \par
\noi October 2002\par \vskip 1 truecm

\noindent e-mails : leyaouan@th.u-psud.fr, oliver@th.u-psud.fr
\newpage
\pagestyle{plain}
Using the OPE and the trace formalism in the heavy quark limit, 
different initial and final four-velocities $v_i$ and $v_f$ for 
initial and final $B$-meson states,
and arbitrary heavy quark currents

\beq
\label{1e}
J_1 = \bar{h}_{v'}^{(c)}\ \Gamma_1\ h_{v_i}^{(b)} \quad , \qquad J_2 
= \bar{h}_{v_f}^{(b)}\ \Gamma_2\ h_{v'}^{(c)}
\eeq

\noi the following sum rule can be written, as shown in ref. \cite{1r}~:

\bea
\label{2e}
&&\Big \{ \sum_{D=P,V} \sum_n Tr \left [ \bar{\cal B}_f(v_f) 
\bar{\Gamma}_2 {\cal D}^{(n)}(v') \right ] Tr \left [ \bar{\cal 
D}^{(n)}(v') \Gamma_1 {\cal
B}_i(v_i)\right ] \xi^{(n)} (w_i) \xi^{(n)} (w_f) \nn \\
&&+ \ \hbox{Other excited states} \Big \}  = - 2 \xi(w_{if})Tr \left 
[ \bar{\cal B}_f(v_f) \bar{\Gamma}_2 P'_+ \Gamma_1 {\cal 
B}_i(v_i)\right ]\ .\eea

In this formula $v'$ is the intermediate meson four-velocity, $P'_+ = 
{1 \over 2} (1 + {/ \hskip - 2 truemm v}')$ comes from the residue of 
the positive energy
part of the $c$-quark propagator, and $\xi(w_{if})$ is the elastic 
Isgur-Wise function that appears because one assumes $v_i \not= v_f$. 
${\cal B}_i$ and
${\cal B}_f$ are the $4 \times 4$ matrices of the ground state $B$ or 
$B^*$ meson and ${\cal D}^{(n)}$ those of all possible ground state 
or excited state $D$
mesons coupled to $B_i$ and $B_f$ through the currents. In formula 
(\ref{2e}) we have made explicit the $j = {1 \over 2}^-$ $D$ and 
$D^*$ mesons and their radial
excitations. \par

The variables $w_i$, $w_f$ and $w_{if}$ are defined as
   \beq \label{3e}
w_i = v_i \cdot v' \qquad w_f = v_f \cdot v' \qquad w_{if} = v_i \cdot v_f \ .
\eeq

The domain of $(w_i$, $w_f$, $w_{if}$) is

\beq
\label{4e}
w_i, w_f \geq 1 \ , \quad w_iw_f - \sqrt{(w_i^2 - 1) (w_f^2 - 1)} 
\leq w_{if} \leq w_iw_f + \sqrt{(w_i^2 -1) ( w_f^2 - 1)} \ .
\eeq

\noi There is a subdomain for $w_i = w_f = w$~:

\beq
\label{5e}
w \geq 1 \ , \qquad 1 \leq w_{if} \leq 2w^2-1 \ .
\eeq

Calling now $L(w_i, w_f, w_{if})$ the l.h.s. and $R(w_i, w_f, 
w_{if})$ the r.h.s. of (\ref{2e}), this SR writes

\beq
\label{6e}
L\left ( w_i, w_f, w_{if} \right ) = R \left ( w_i, w_f, w_{if} \right )
\eeq

\noi where $L(w_i, w_f, w_{if})$ is the sum over the intermediate $D$ 
states and $R(w_i, w_f, w_{if})$ the OPE side. Within the domain 
(\ref{4e}) one can derive
relatively to any of the variables $w_i$, $w_f$ and $w_{if}$

\beq
\label{7e}
{\partial^{p+q+r} L \over \partial w_i^p \partial w_f^q \partial 
w_{if}^r} = {\partial^{p+q+r} R \over \partial w_i^p \partial w_f^q 
\partial w_{if}^r}
\eeq
\vskip 3 truemm

\noi and obtain different SR taking different limits to the frontiers 
of the domain, e.g.
\bea
\label{8e}
& &\qquad w_{if} \to 1, w_i = w_f = w \nn \\
&\hbox{or} &\qquad w_i \to 1, w_{if} = w_f = w\nn \\
&\hbox{or} &\qquad w_f \to 1, w_{if} = w_i = w \ .\eea

The aim of the present paper is to present some results on the 
derivatives of the elastic Isgur-Wise function $\xi (w)$ at zero 
recoil $w = 1$.\par

As in ref. \cite{1r}, we choose as initial and final states the $B$ meson,
\beq
\label{9e}
{\cal B}_i (v_i) = P_{i+} (- \gamma_5) \qquad {\cal B}_f (v_f) = 
P_{f+} (- \gamma_5)
\eeq

\noi and vector or axial currents projected along the $v_i$ and $v_f$ 
four-velocities. Choosing the vector currents

\beq
\label{10e}
J_1 = \bar{h}_{v'}^{(c)}\ {/ \hskip - 2 truemm v}_i\ h_{v_i}^{(b)} 
\quad , \qquad J_2 = \bar{h}_{v_f}^{(b)}\ {/ \hskip - 2 truemm v}_f\ 
h_{v'}^{(c)} \ .
\eeq

\noi and gathering the formulas (60) and (89)-(91) of ref. \cite{1r} we 
obtain the SR (\ref{2e}) with the sum of all excited states $j^P$ in 
a compact form~:
\bea
\label{11e}
&&(w_i + 1) (w_f + 1) \sum_{\ell \geq 0} {\ell + 1 \over 2 \ell + 1} 
S_{\ell} (w_i, w_f, w_{if}) \sum_n \tau_{\ell + 1/2}^{(\ell)(n)}(w_i)
\tau_{\ell + 1/2}^{(\ell )(n)}(w_f)\nn \\
&&+ \sum_{\ell \geq 1} S_{\ell} (w_i, w_f, w_{if}) \sum_n \tau_{\ell 
- 1/2}^{(\ell)(n)}(w_i)
\tau_{\ell - 1/2}^{(\ell )(n)}(w_f)\nn \\
&&= (1 + w_i+w_f+w_{if}) \xi(w_{if}) \ .\eea

Choosing instead the axial currents

\beq
\label{12e}
J_1 = \bar{h}_{v'}^{(c)}\ {/ \hskip - 2 truemm v}_i\ \gamma_5 \ 
h_{v_i}^{(b)} \quad , \qquad J_2 = \bar{h}_{v_f}^{(b)}\ {/ \hskip - 2 
truemm v}_f\
\gamma_5\ h_{v'}^{(c)}
\eeq

\noi the SR (\ref{2e}) writes, from the formulas (48) and (92)-(94) 
of ref. \cite{1r}~:
\bea
\label{13e}
&& \sum_{\ell \geq 0} S_{\ell + 1} (w_i, w_f, w_{if}) \sum_n 
\tau_{\ell + 1/2}^{(\ell)(n)}(w_i)
\tau_{\ell + 1/2}^{(\ell )(n)}(w_f)\nn \\
&&+ (w_i - 1) (w_f - 1) \sum_{\ell \geq 1} {\ell \over 2 \ell - 1} 
S_{\ell - 1} (w_i, w_f, w_{if}) \sum_n \tau_{\ell - 
1/2}^{(\ell)(n)}(w_i) \tau_{\ell - 1/2}^{(\ell
)(n)}(w_f) \nn \\
&&= - (1 - w_i-w_f+w_{if}) \xi (w_{if}) \ .\eea

Following the formulation of heavy-light states for arbitrary $j^P$ 
given by Falk \cite{2r}, we have defined in ref. \cite{1r} the IW 
functions $\tau_{\ell +
1/2}^{(\ell)(n)}(w)$ and $\tau_{\ell - 1/2}^{(\ell)(n)}(w)$, that 
correspond to the orbital angular momentum $\ell$ of the light quark 
relative to the heavy
quark, $j = \ell \pm {1 \over 2}$ being the total angular momentum of 
the light cloud. We could made explicit the contribution of
the states ${1 \over 2}^-$, ${1 \over 2}^+$, ${3 \over 2}^+$, using 
the traditional notation of Isgur and Wise \cite{3r}. For the lowest 
values of $\ell$, one has
the identities

\beq
\label{14e}
\tau_{1/2}^{(0)}(w) \equiv \xi (w) \quad , \quad \tau_{1/2}^{(1)}(w) 
\equiv 2 \tau_{1/2}(w) \quad , \quad \tau_{3/2}^{(1)}(w) \equiv 
\sqrt{3}\  \tau_{3/2}(w)
\eeq

\noi where a radial quantum number is implicit. Therefore, the 
functions $\tau_{1/2}^{(1)}(w)$ and $\tau_{3/2}^{(1)}(w)$ correspond, 
respectively, to the functions
$\zeta (w)$ and $\tau (w)$ defined by Leibovich et al. \cite{4r}. But 
in all this paper we will keep the general notation $\tau_{\ell + 
1/2}^{(\ell)(n)}(w)$,
$\tau_{\ell - 1/2}^{(\ell)(n)}(w)$.  \par

In equations (\ref{10e}) and (\ref{12e}) the quantity $S_n$ is defined by

\beq
\label{15e}
S_n = v_{i\nu_1} \cdots v_{i\nu_n}\ v_{f\mu_1} \cdots v_{f\mu_n} \ T^{\nu_1 \cdots \nu_k, \mu_1 \cdots \mu_n}
\eeq
\noi and the polarisation tensor $T^{\nu_1 \cdots \nu_k, \mu_1 \cdots \mu_n}$, given by
 
\beq
\label{16e}
T^{\nu_1 \cdots \nu_k, \mu_1 \cdots \mu_n} = \sum_{\lambda} 
\varepsilon'^{(\lambda )*\nu_1 \cdots \nu_n} \ \varepsilon'^{(\lambda 
)\mu_1 \cdots \mu_n}
\eeq

\noi depends only on the four-velocity $v'$. The polarization tensor 
$\varepsilon '^{\mu_1 \cdots \mu_n}$ is a traceless symmetric tensor, 
i.e. a symmetric tensor
with vanishing contractions. Moreover, as demonstrated in the 
Appendix A of ref. \cite{1r}, $S_n$ is given by the following 
expression~:

\beq
\label{17e}
S_n(w_i,w_f,w_{if}) = \sum_{0 \leq k \leq {n \over 2}} C_{n,k} (w_i^2 
- 1)^k (w_f^2 - 1)^k (w_i w_f - w_{if})^{n-2k}
\eeq

\noi with

\beq
\label{18e}
C_{n,k} = (-1)^k {(n!)^2 \over (2n) !} \ {(2n - 2k) ! \over k! (n-k) 
! (n-2k)!} \ .
\eeq

Making the sum of both equations (\ref{11e}) and (\ref{13e}), one obtains

\vskip 5 truemm
\bea
\label{19e}
&&2(w_i + w_f) \xi (w_{if}) \nn \\
&&= \sum_{\ell \geq 0} \left [ {\ell + 1\over 2 \ell + 1} (w_i + 1) 
(w_f + 1) S_{\ell} (w_i, w_f , w_{if}) + S_{\ell + 1} (w_i, w_f , 
w_{if}) \right ] \nn \\
&&\sum_n \tau_{\ell + 1/2}^{(\ell )(n)}(w_i) \tau_{\ell + 1/2}^{(\ell 
)(n)}(w_f)\nn \\
&&+ \sum_{\ell \geq 1} \left [ S_{\ell} (w_i, w_f , w_{if}) + {\ell 
\over 2 \ell - 1} (w_i - 1) (w_f - 1) S_{\ell - 1} (w_i, w_f , 
w_{if}) \right ] \nn \\
&&\sum_n \tau_{\ell - 1/2}^{(\ell)(n)}(w_i)  \tau_{\ell - 1/2}^{(\ell 
)(n)}(w_f) \ .
\eea

Taking the $m$-th derivative of this relation relatively to $w_{if}$ 
and using

\beq
\label{20}
\left. {\partial^m \over \partial w_{if}^m} \ S_n (w_i, w_f w_{if}) 
\right |_{w_i=w_f=w_{if} = 1} = (-1)^n \ n ! \ \delta_{m,n}
\eeq

\noi one readily obtains the derivatives at zero recoil
  \bea
\label{21e}
&&\xi^{(\ell )} (1) = {1 \over 4} \ (-1)^{\ell} \ \ell ! \left \{ 
{\ell + 1 \over 2 \ell + 1} 4 \sum_n \left [ \tau_{\ell + 1/2}^{(\ell 
)(n)}(1) \right
]^2 \right . \nn \\
&&\left . + \sum_n \left [ \tau_{\ell - 1/2}^{(\ell -1)(n)}(1) \right 
]^2 + \sum_n \left [ \tau_{\ell - 1/2}^{(\ell)(n)}(1) \right ]^2 
\right \} \quad (\ell \geq
0) \ .\eea

\noi This relation shows that $\xi (w)$ is an alternate series in 
powers of $(w-1)$. Equation (\ref{21e}) gives 1 = 1 for $\ell = 0$, 
and reduces to Bjorken SR
\cite{5r}, \cite{3r} for $\ell = 1$ (with the notation of Isgur and 
Wise (\ref{14e}))~:

\beq
\label{22e}
\rho^2 = {1 \over 4} + \sum_n \left [ \tau_{1/2}^{(n)}(1) \right ]^2 
+ 2 \sum_n \left [ \tau_{3/2}^{(n)}(1) \right ]^2 \ .
\eeq

\vskip 3 truemm
\noi We consider now the SR obtained from the axial current 
(\ref{13e}), taking the $m$-th derivative relatively to $w_{if}$ and 
making $w_i=w_f = w_{if} = 1$ one
obtains from (\ref{20}) the simple formula~: \beq
\label{25e}
\xi^{(\ell )} (1) = \ell !\ (-1)^{\ell} \sum_n \left [ \tau_{\ell + 
1/2}^{(\ell)(n)}(1) \right ]^2 \quad (\ell \geq
0) \ .
\eeq

\noi This expression is the generalization to the $\ell$-th 
derivative of the relation obtained in ref. \cite{1r} for the slope 
(with the notation (\ref{14e}),
formula (55))~:

\beq
\label{26e}
\rho^2 = 3 \sum_n \left [ \tau_{3/2}^{(n)}(1) \right ]^2 \ .
\eeq

\noi Combining (\ref{21e}) and (\ref{25e}) one obtains the relation 
among the IW functions~:

\beq
\label{27e}
{\ell \over 2 \ell + 1} \sum_n \left [ \tau_{\ell + 
1/2}^{(\ell)(n)}(1) \right ]^2 - {1 \over 4}  \sum_n \left [ 
\tau_{\ell - 1/2}^{(\ell)(n)}(1) \right ]^2 =
{1 \over 4}  \sum_n \left [ \tau_{\ell - 1/2}^{(\ell - 1)(n)}(1) 
\right ]^2   \eeq

\noi that gives $0 = 0$ for $\ell = 0$ and reduces to Uraltsev SR 
\cite{6r} for $\ell = 1$ (with the notation (\ref{14e}))~:

\beq
\label{28e}
\sum_n \left [ \tau_{3/2}^{(n)}(1) \right ]^2 -  \sum_n \left [ 
\tau_{1/2}^{(n)}(1) \right ]^2 = {1 \over 4}
\eeq

\noi and generalizes it for all $\ell$. \par

Replacing now $\sum\limits_n [ \tau_{\ell + 1/2}^{(\ell)(n)}(1) ]^2$ 
from the expression (\ref{25e}) into the generalization of Bjorken SR 
(\ref{21e}) one obtains~:

\beq
\label{29e}
(-1)^{\ell} \ \xi^{(\ell)} (1) = {1 \over 4} \ {2 \ell + 1 \over 
\ell} \ell ! \left \{ \sum_n \left [ \tau_{\ell - 1/2}^{(\ell - 
1)(n)}(1) \right ]^2 + \sum_n \left [ \tau_{\ell -
1/2}^{(\ell)(n)}(1) \right ]^2 \right \} \ .
  \eeq

\vskip 3 truemm
\noi For $\ell = 1$ one obtains~:

\beq
\label{30e}
\rho^2 = - \xi '(1) = {3 \over 4} \left \{ 1 + \sum_n \left [ 
\tau_{1/2}^{(1)(n)}(1) \right ]^2 \right \}
\eeq

\noi and therefore~:

\beq
\label{31e}
\rho^2 \geq  {3 \over 4}
\eeq

\noi i.e., one obtains the lower bound for the slope obtained from 
the combination of Bjorken and Uraltsev SR (\ref{22e}) and 
(\ref{28e}). \par

For $\ell = 2$ one gets~:

  \beq
\label{32e}
\xi '' (1) = {1 \over 4} \ {5 \over 2}\ 2! \left \{ \sum_n \left [ 
\tau_{3/2}^{(1)(n)}(1) \right ]^2 + \sum_n \left [ 
\tau_{3/2}^{(2)(n)}(1) \right ]^2 \right \}
\eeq

\noi and

\beq
\label{33e}
\sigma^2 = \xi '' (1) \geq {1 \over 4} \ {5 \over 2}\ 2! \ \sum_n 
\left [ \tau_{3/2}^{(1)(n)}(1) \right ]^2 \ .
\eeq

\noi Using equation (\ref{26e}), with the notation (\ref{14e}), we 
have therefore demonstrated on rigorous grounds the bound proposed, 
modulo a phenomenological
hypothesis, in ref. \cite{1r}~:

\beq
\label{34e}
\sigma^2 \geq {5 \over 4} \rho^2
\eeq

\noi that implies, from (\ref{31e}), the absolute bound

\beq
\label{35e}
\sigma^2 \geq {15 \over 16} \ .
\eeq

\noi One can similarly bound any derivative $(-1)^{\ell} 
\xi^{(\ell)}(1)$ in terms of the precedent one $(-1)^{\ell - 1} 
\xi^{(\ell - 1)}(1)$~:

\beq
\label{36e}
(-1)^{\ell} \xi^{(\ell)}(1) \geq  {2\ell + 1 \over 4} \left [ 
(-1)^{\ell - 1} \xi^{(\ell - 1)}(1) \right ]
\eeq

\noi implying the following absolute bound for the $\ell$-th 
derivative of the Isgur-Wise function~:

  \beq \label{37e}
(-1)^{\ell} \xi^{(\ell)}(1) \geq  {(2\ell + 1)!!  \over 2^{2\ell}}
\eeq

\noi that gives, in particular, (\ref{31e}) and (\ref{35e}) for the 
lower cases. Notice that this bound is satisfied by the 
phenomenological dipole formula

\bea
\label{38e}
\xi_{Dipole}(w) = \left ( {2 \over w+1}\right )^2
\eea

\noi obtained approximately within the Bakamjian-Thomas relativistic 
quark models \cite{7r} and from QCD Sum Rules at large recoil 
\cite{8r}. Indeed, one can
easily show that the derivatives

  \beq
\label{39e}
(-1)^{\ell} \ \xi_{Dipole}^{(\ell)}(1) = {(\ell + 1)!  \over 2^{\ell}}
\eeq

\noi satisfy (\ref{36e}) and (\ref{37e}). \par

An interesting phenomenological remark is that a parametrization of 
the IW function of the form
\beq
\label{40e}
\xi (w) = \left ( {2 \over w + 1}\right )^{2\rho^2}
\eeq

\noi with $\rho^2 \geq {3 \over 4}$ satisfies also the inequalities 
(\ref{36e}) and (\ref{37e}). At the lower bound $\rho^2 = {3 \over 
4}$, (\ref{40e})
satisfies (\ref{36e}) and (\ref{37e}) with the equality.\par

The result (\ref{37e}), that shows that all derivatives at zero 
recoil are large, should have important
phenomenological implications for the empirical fit needed for the 
extraction of $|V_{cb}|$ in $B \to D^*\ell \nu$ \cite{9r}. In the 
case of the slope and the
curvature, the lower bounds (\ref{31e}) and (\ref{35e}) are 
complementary to the upper bounds obtained from unitarity constraints 
\cite{10r}, \cite{11r}.\\

\noi {\large \bf Acknowledgements} \par
We are indebted to Nikolai Uraltsev for long standing
discussions on QCD in the heavy quark limit. We acknowledge 
support from the EC contract HPRN-CT-2002-00311 (EURIDICE).

\end{document}